\documentclass[a4paper,fleqn,usenatbib]{mnras}

\usepackage{newtxtext,newtxmath}
\usepackage[T1]{fontenc}
\usepackage{ae,aecompl}

\usepackage{graphicx}	% Including figure files
\usepackage{amsmath}	% Advanced maths commands
\usepackage{amssymb}	% Extra maths symbols
\usepackage{url}
\usepackage{hyperref}
\hypersetup{colorlinks=false,pdfborder={0 0 0}}
\usepackage{etoolbox}\usepackage[space]{grffile}
\usepackage{latexsym}
\usepackage{textcomp}
\usepackage{longtable}
\usepackage{tabulary}
\usepackage{booktabs,array,multirow}
\usepackage{float}

\title[T$_\mathrm{eff}$ for exoplanet direct imaging]{Toward the Analysis of JWST Exoplanet Spectra: the effective temperature in the context of direct imaging}

\author[J.-L. Baudino et al.]{
Jean-Loup Baudino,$^{1}$\thanks{E-mail: dr.jean-loup.baudino@hotmail.com}
J. Taylor,$^{1}$
P. G. J. Irwin$^{1}$
and R. Garland$^{1}$
\\
$^{1}$ Department of Physics, University of Oxford, Oxford OX1 3PU, UK
}

\date{Accepted 2019 September 28. Received 2019 September 27; in original form 2019 July 9}

\pubyear{2019}

\begin{document}
\label{firstpage}
\pagerange{\pageref{firstpage}--\pageref{lastpage}}
\maketitle

% Abstract of the paper 250 words
\begin{abstract}
The current sparse wavelength range coverage of exoplanet direct imaging observations, and 
the fact that models are defined using a finite wavelength range, lead both to uncertainties 
on effective temperature determination. We study these effects using black-bodies and 
atmospheric models and we detail how to infer this parameter. Through highlighting the key 
wavelength coverage that allows for a more accurate representation of the effective 
temperature, our analysis can be used to mitigate or manage extra uncertainties being added 
in the analysis from the models. We find that the wavelength range coverage will soon no 
longer be a problem. An effective temperature computed by integrating the spectroscopic 
observations of the James Webb Space Telescope (JWST) will give uncertainties similar to, or 
better than, the current state--of--the--art, which is to fit models to data.  Accurately 
calculating the effective temperature will help to improve current modelling approaches. 
Obtaining an independent and precise estimation of this crucial parameter will help the 
benchmarking process to identify the best practice to model exoplanet atmospheres.
\end{abstract}


%%%%%%%%%%%%%%%%%%%%%%%%%%%%%%%%%%%%%%%%%%%%%%%%%%

%%%%%%%%%%%%%%%%% BODY OF PAPER %%%%%%%%%%%%%%%%%%

\section{Introduction}
\label{sec-1}

Since the first confirmation of a directly detected exoplanet by \cite{Chauvin_2005}, the 
effective temperature ($T_\mathrm{eff}$) has been one of the main parameters analysed. 
Exoplanet direct imaging gives planetary astronomers the opportunity to characterise the 
atmospheres of young giant exoplanets using their emission spectra obtained with ground-based 
instruments such as the Spectro-Polarimetric High-contrast Exoplanet REsearch 
\citep[SPHERE][]{Beuzit_2008, Beuzit_2019}, the Gemini Planet 
Imager~\citep[GPI][]{Macintosh_2014} or soon from space with the Near-Infrared Spectrograph 
(NIRSpec) and the Mid-Infrared Instrument~\citep[MIRI,][]{Rieke_2015,Wright_2015} of the 
JWST. The current state--of--the--art of instrumentation probes the outer part ($\geq$~10~AU) 
of young ($<$~100~Myrs) stellar systems, studying young exoplanets with spectra overly 
dominated by self-emission coming from formation and contraction processes.

The temperature of an astrophysical object can be defined by the effective 
temperature~\citep{Hopkins_1864},  it is theoretically defined by using the black-body 
theory~\citep{Violle_1892, LeChatelier_1892, Wilson_1894}:
\begin{equation}
   M = \int_0^{+\infty} F_{\lambda} d\lambda =\sigma T_\mathrm{eff}^4
   \label{eq-blackbody}
\end{equation}
where \textit{M} is the bolometric flux per unit surface area, $F_{\lambda}$ is the emission flux density in 
$W m^{-2} \mu m^{-1}$, and $\sigma$ the Stefan--Boltzmann constant.\\

There are three techniques that are used to determine the effective temperature of an 
exoplanet. Generally the community fits the data with models~\citep[e.g.][]{Bonnefoy_2016}, 
mainly to deal with the sparse wavelength coverage of real observations, for the majority of 
the exoplanets, in the Near Infrared (NIR, between $J$ and $L$ bands, i.e. 
$\sim$~1.2--4~$\mu$m). The models used are also defined on a finite wavelength range, so that 
the integration is defined over a limited wavelength range and not between 0 and 
$+\infty~\mu$m as in the Eq.~\ref{eq-blackbody}. This situation can introduce an intrinsic 
error on the estimated $T_\mathrm{eff}$.

Retrieval approaches struggle to determine $T_\mathrm{eff}$, intrinsically because this 
technique computes the spectrum of the atmosphere only where the data is available and 
instead will often focus on retrieving a temperature profile that can be difficult to link to 
$T_\mathrm{eff}$.

Another approach for estimating $T_\mathrm{eff}$ was proposed by \cite{Morzinski_2015}, who 
extrapolated directly the bolometric flux from the data for the planet $\beta$~Pictoris~b 
(assuming black-body emission spectra). As we show later in this paper, this approach was 
successful thanks to an exceptional data quality and coverage, and we will look ahead to what 
we will be able to do with JWST.\\

The JWST is the next generation of space-based international observatory. It is planned to be 
launched in early 2021, and will be game-changing in terms of exoplanet characterisation, 
especially for directly-imaged exoplanets \citep{Danielski_2018}. As soon as the observations 
begin, by combining observations from two instruments (NIRSpec and MIRI) with a resolution 
and signal-to-noise ratio (SNR) never obtained before, the community (Early Release Science 
programme ID 1386, PI Sasha Hinkley) plans to obtain the first complete measurement of the 
spectral energy distribution (SED) of an exoplanet. With a SNR$>$100, the spectra provided by 
JWST will have better quality than the current state--of--the--art of 
modelling~\citep{Baudino_2017} and will cover the major part of the emission flux of the 
planets.\\  

The this paper give a review of the current technique fo inferring $T_\mathrm{eff}$ from 
observations to explore the weaknesses and help modellers manage the intrinsic errors 
linked to these methods. We then highlight and quantify one of the opportunities given to the 
community by the JWST to obtain a rigorous estimate of $T_\mathrm{eff}$ directly from the 
observations and the impact of this on the future of exoplanet atmosphere modelling.\\

We begin this paper by exploring the effect of wavelength range completeness 
(Sec.~\ref{sec-2}) using simple black-body functions and synthetic spectra from 
Exo-REM~\citep[a radiative-convective atmospheric model of exoplanets][]{Baudino_2015,Baudino_2017} 
and two other models from the literature (Ydwarf model from \citealt{Morley_2012} and 
Drift-PHOENIX,~\citealt{Helling_2008}). We follow by reviewing how the effective temperature 
is inferred, including a proposal of how it can be achieved using retrieval methods and a way 
to evaluate this parameter without models (Sec.~\ref{sec-3}). We then quantify the 
uncertainties on the temperature using the future observations of the JWST (Sec.~\ref{sec-4}) 
and we explain how this may be used to improve exoplanet modelling. Finally we summarise our 
conclusions (Sec.~\ref{sec-5}).

\section{Completeness effect}
\label{sec-2}

In this section, we explore in depth the calculation of $T_\mathrm{eff}$. First, 
we compare pure black-body spectra to more realistic spectra generated by self-consistent 
models with various wavelength range coverages. We also show that we can use black-body 
spectra as a proxy for models.

\subsection{Black-body}
\label{sec-2-1}

Our first approach is to explore the effect of estimating $T_\mathrm{eff}$ from an incomplete 
wavelength range for the case of simple black-body emission. We use black-body spectra for a 
range of $T_\mathrm{eff}$ between 400 and 1800~K with step size of 200~K; for each case 
we integrate the spectrum between a minimum of 0.3 $\mu$m and a maximum $\lambda_{cut}$ 
between 5 and 500~$\mu$m in $M(T_\mathrm{eff},\lambda_{cut})$.

Then we study how ${M((T_\mathrm{eff},\lambda_{cut})/\sigma)^{1/4}}$ diverges compared to 
$M((T_\mathrm{eff},500~\mu m)/\sigma)^{1/4}$ using the resultant absolute difference in 
temperature $\delta T$:

\begin{equation}
   \delta T = T_\mathrm{eff}* 
   \frac{M((T_\mathrm{eff},\lambda_{cut})/\sigma)^{1/4}-
   M((T_\mathrm{eff},500~\mu m)/\sigma)^{1/4}}{M((T_\mathrm{eff},500~\mu m)/\sigma)^{1/4}}
	\label{eq-deltateff}
\end{equation}

We approximate the $\delta T$ as an intrinsic uncertainty of the $T_\mathrm{eff}$ resulting 
from limiting the wavelength range.

We also performed tests on the minimum (0.3 $\mu$m) and maximum (500$\mu$m) wavelength 
boundaries and expanding the wavelength range (for example from 0.1--1000 $\mu$m) did not 
significantly impact our results.

\subsection{Atmospheric models}
\label{sec-2-2}

To test our analysis on more realistic cases, we make use of six models spanning the 
temperature range. With Exo-REM we generate two test cases with $T_\mathrm{eff} =$ 500 and 
1500~K. We also complete the comparison by selecting two models from the literature with two 
different $T_\mathrm{eff}$: 200~K and 450~K for the Ydwarf model and 1000~K and 1800~K for 
Drift-PHOENIX. Following the method described in Sec.~\ref{sec-2-1}, we compare our 
black-body cases with these test cases. The data, except for Exo-REM, was taken from the VOSA 
Theoretical spectra web server\footnote{\url{http://svo2.cab.inta-csic.es/theory/newov2/}} \citep{Bayo_2008}.

\subsection{Theoretical effect}
\label{sec-2-3}

\begin{figure}
\begin{center}
\includegraphics[width=1.00\columnwidth]{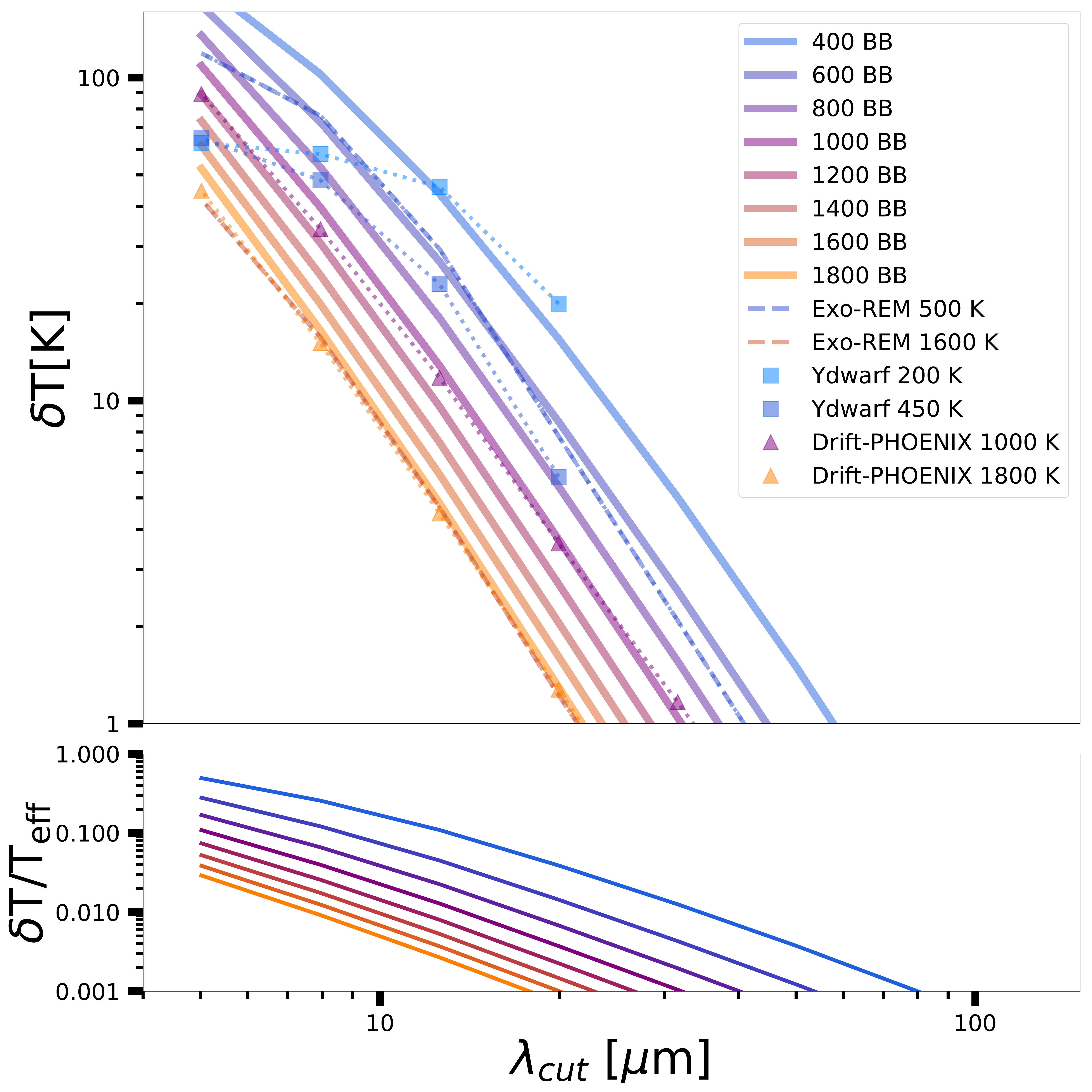}
\caption{Top panel: The differences ($\delta$T) between the $T_\mathrm{eff}$ calculated from 0.3 $\mu$m 
to $\lambda_{cut}$, compared with an integration between 0.3 and 500~$\mu$m 
(eq.~\ref{eq-deltateff}), depending of the wavelength where the integration is stopped, 
$\lambda_{cut}$. The colour code is linked to the temperature. Plain curves are the 
black-body emissions (BB). The dashed lines are for Exo-REM. The dotted lines with squares 
are for Ydwarf. The dotted with triangles are for Drift-PHOENIX. Bottom panel: Ratio between $\delta$T 
and $T_\mathrm{eff}$ compared with $\lambda_{cut}$. The color code is identical to the top panel.}
\label{Teffvsrange}
\end{center}
\end{figure}

To begin the analysis, we first compare the relationship between the uncertainties on the 
final estimated $T_\mathrm{eff}$ and the completeness of wavelength coverage of the model 
used ("model" is used here as a generic word for either black-body or more complex models). 
Fig.~\ref{Teffvsrange} shows the uncertainties on the $T_\mathrm{eff}$, $\delta T$, compared 
to the $\lambda_{cut}$ for all the black-body spectra and the model test cases with a colour 
code linked to the temperature. 

The first visible effect is the fact that the lower $T_\mathrm{eff}$ will be mostly 
significantly impacted by the wavelength range completeness. This is easy to explain by the 
fact that the maximum of emission is, by definition, more in the red part (toward the longer 
wavelengths) of the spectrum for the cold cases, so with a maximum near $\lambda_{cut}$ and 
consequently with more flux missing.

Secondly the selection of the $\lambda_{cut}$ impacts the $\delta T$ monotonically and 
logarithmically due to the fact that the maximum of the flux is always included in the 
wavelength range ($\sim$~7~$\mu$m for 400~K) and we are just completing the integration with 
the remaining part of the tail of the black-body function. For the worst case shown 
($T_\mathrm{eff}=$400~K) a $\lambda_{cut}>$70~$\mu$m gives a $\delta T<$1~K. 

If we focus on non-negligible uncertainties ($\delta T >$~1 K), model cases are similar or 
better in $\delta T$ compared to their black-body analogues. This effect can be explained by 
the fact that, generally, models generate spectra with the vast majority of the energy 
concentrated below 10~$\mu$m. Hence, for the next section we can use the black-body 
approximation for the shape of the spectra, this is representative of the worst case scenario.

\section{Evaluation of the temperature}
\label{sec-3}

In this section we detail how observations can be used to infer the $T_\mathrm{eff}$ 
using self-consistent models, retrieval or by directly using the data. Hence, there are three 
ways to infer $T_\mathrm{eff}$.\\

\subsection{Self-consistent models}
\label{sec-3-1}

For self-consistent models, the $T_\mathrm{eff}$ is an input parameter used to generate a 
spectrum. The data are compared to a set of spectra: the subset of spectra that best reproduce 
the data give the $T_\mathrm{eff}$ of the observed target 
\citep[e.g.][]{Macintosh_2015, Bonnefoy_2018}. In this approach the uncertainty coming from 
the completeness of the wavelength range is not obvious to manage. For example in Exo-REM the 
spectrum is computed between 0.65 and 500~$\mu$m: according to the previous section the 
impact is negligible in this case. But each model has its own definition and the Drift-
PHOENIX and Ydwarf spectra are available in the VOSA database between 0.0001 and 950~$\mu$m 
and 0.4 and 50~$\mu$m respectively. This last range could be problematic if that was the 
range over which the structure of the atmosphere was computed, but this is only the published 
version of the spectra, as \cite{Morley_2012} refers (Sec.~2.3) to a wavelength range of 
0.268 to 227~$\mu$m for the model computation itself.

The $T_\mathrm{eff}$ and its relative uncertainty is normally derived 
by combining the results of various models from various teams. This is done to prevent the 
result being biased by the hypothesis of one model. The range of $T_\mathrm{eff}$ will either 
be defined by the "best" result for each model \citep[e.g.][]{Bonnefoy_2018} or by combining 
the range of models reproducing the data at a given level of uncertainty 
\citep[e.g.][]{Baudino_2015}. The "best" result is normally determined using the $\chi^2$ or 
the $G$ goodness-of-fit parameter \citep{Cushing_2008} and some models can be discarded in the 
process if they don't reach a given threshold. Instead of keeping an unique solution for each 
model it is also possible to keep the set of solutions for each model where the $\chi^2$ is 
linked to a given confidence interval (e.g. 1, 2, 3~$\sigma$).

The typical uncertainty quoted in the literature is often around 100 to 200~K for an 
inferred temperature range between 500--2000~K (with many objects with a $T_\mathrm{eff}$ 
around 1000~K).\\

The main bias arising from self-consistent models comes from the physics that has been 
assumed, but self-consistent model results are also impacted by our knowledge in terms of 
theoretical spectroscopy \citep{Baudino_2017}. However, following the improvement of the data 
quality (increased signal-to-noise ratio, resolution, wavelength coverage), we can increase 
the complexity our models, hence the number of free parameters. One convenient method to use 
in this case, coming from Earth and Solar System studies, is to use a retrieval approach 
\citep[e.g.][]{Irwin_2008} that is not biased by the physics considered.

\subsection{Retrieval}
\label{sec-3-2}

Defining the $T_\mathrm{eff}$ using a retrieval technique is more difficult. Driven by the 
observations, retrieval techniques intrinsically do not generate the full spectrum and, by 
doing so, cannot determine this parameter. 

The temperature is taken into account by retrieving the temperature profile where often one 
parameter can be similar to the $T_\mathrm{eff}$ \citep[e.g. $T_\mathrm{int}$ in][is similar 
for three of the four planets of the HR~8799 system]{Lavie_2017}, but this parameter is 
difficult to compare with self-consistent models. The other possibility is to consider a 
given atmosphere (e.g. grey) and so the $T_\mathrm{eff}$ can be derived as a simple 
parameter of the temperature profile \citep[e.g.][sec. 6.2.1]{Garland_2018}. However, this 
last approach leads to a difference in the $T_\mathrm{eff}$ compared to self-consistent 
models; for example, we observed a difference of 200--300~K higher, by applying for GJ~504~b 
\citep[test done to prepare][]{Bonnefoy_2018} the profile defined in \cite{Garland_2018}.\\

If the observations are good enough and include a wavelength range with molecular features, 
another approach that we are proposing is to retrieve at least the temperature profile and 
molecular abundances of molecules with global impact (mainly H$_2$O). We can then use this 
result to define an atmospheric structure and generate a full spectrum of the object that can 
be used to compute $T_\mathrm{eff}$. This last approach does not retrieve the parameter 
directly, but gives the temperature for the retrieved atmosphere and gives similar results 
compared to self-consistent models. This will be explored in a later study.\\

Even though retrievals are "data-driven", there are still potential biases arising from model 
assumptions, similar to those of self-consistent models, such as how the radiative transfer 
calculation is actually computed (e.g. assumptions of line shape, sub-Lorentzian correction, 
etc.). In the next section we show how we can try to avoid that, by using directly the 
observational data to compute the $T_\mathrm{eff}$.

\subsection{Extrapolation from the observation}
\label{sec-3-3}

Occasionally, the observational data cover a sufficient wavelength range to enable us to 
attempt to derive $T_\mathrm{eff}$ by extrapolating it from the observation 
\citep{Morzinski_2015}. $\beta$~Pictoris~$b$, the planet studied by \cite{Morzinski_2015}, is 
one of the most studied directly-imaged objects and the large wavelength coverage available 
(0.9--4.5~$\mu$m) is pretty unique. By integrating the observed flux and completing it by a 
black-body approximation outside the observed wavelength range to cover 0.001 to 
100~$\mu$m, and by evaluating the error bars using a Monte Carlo experiment 
\citep[Sec.~4.4 in][]{Morzinski_2015}, this study finds a $T_\mathrm{eff}$ consistent with 
the comparison of the data \citep[Tab.~15 in][]{Morzinski_2015} with self-consistent models 
and shows a first step towards an independent determination of this temperature. 

The biggest problem with applying this method is the current need to combine various 
instruments (e.g. NaCo, MagAO and GPI) from various observatories (e.g. Very Large Telescope, 
Magellan, Gemini South) that are not well cross-calibrated and with the flux extracted using 
various techniques (e.g. Angular Differential Imaging, Principal Component Analysis, Spectral 
Deconvolution) to populate the wavelength range.\\

It is within this context that we show, in the following section, how JWST will change the 
way how we derive $T_\mathrm{eff}$.

\section{Using observations of the JWST}
\label{sec-4}

\begin{figure}
\begin{center}
\includegraphics[width=1.00\columnwidth]{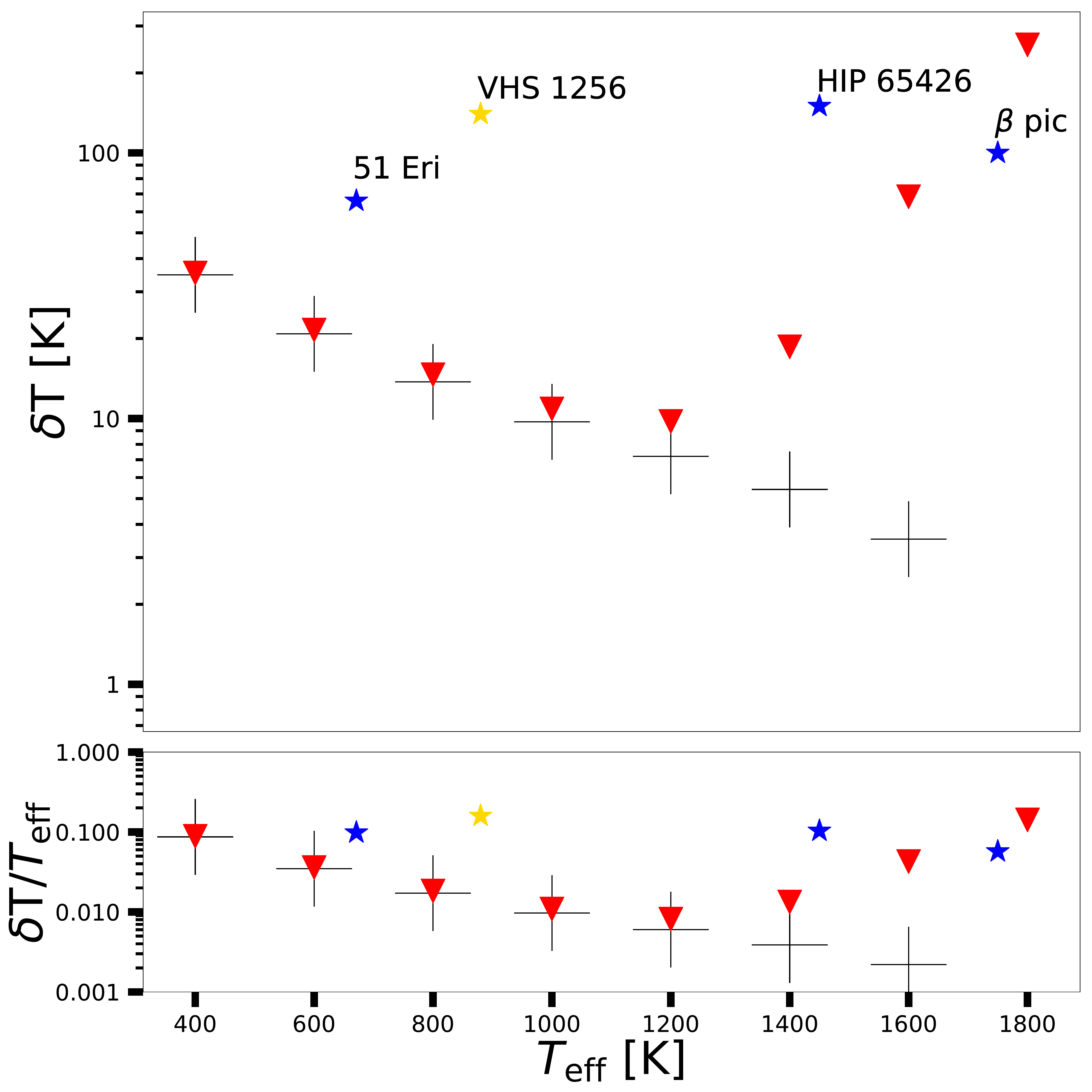}
\caption{Top panel: The uncertainties ($\delta$T) relative to the $T_\mathrm{eff}$ 
calculated for a black-body spectra covering only the NIRSpec and MIRI wavelength ranges 
(black crosses and red triangles), without the consideration of radius uncertainty. The black 
crosses take into account only the effect of wavelength range completeness. The red triangles 
additionally take into account the uncertainties expected for VHS~1256-1257~b (expected to 
be one of the best targets). For the comparison with the current state--of--the--art, the 
stars indicate the estimated uncertainties of $T_\mathrm{eff}$ for four observed exoplanets, 
already published, with ground-based observations: 51~Eridani~b, VHS~1256-1257~b, HIP~65426~b 
and $\beta$~Pictoris~b, characterised using comparison with models. Bottom panel: 
Ratio between $\delta$T and $T_\mathrm{eff}$ relative to the $T_\mathrm{eff}$.
}
\label{dTeff_Teff}
\end{center}
\end{figure}

In this section we analyse the uncertainties on $T_\mathrm{eff}$ if this parameter is defined 
observationally using JWST spectroscopy, without taking into-account the uncertainty 
on the radius. Fig.\ref{dTeff_Teff} (black crosses) shows the error 
on $T_\mathrm{eff}$ using the JWST NIRSpec and MIRI spectral capabilities, i.e. with a 
spectrum covering 0.6--13~$\mu$m. The uncertainties derived directly are expected to be low, 
($< 10~\%$ for the worst case). This gives us the opportunity to obtain $T_\mathrm{eff}$ without 
modelling biases and with an accuracy greater than the current state--of--art using models 
illustrated in the figure of, respectively: 51~Eridani~b~\citep{Rajan_2017}, 
VHS~1256-1257~b~\citep{Gauza_2015}, HIP~65426~b~\citep{Chauvin_2017} and 
$\beta$~Pictoris~b~\citep{Chilcote_2017}.\\ 

Using the SNR expected for VHS~1256-1257~b \citep[i.e. observation of half an hour, with Prism 
and LRS mode, see][]{Baudino_2017}, we also generated the error on $T_\mathrm{eff}$ taking 
into account favorable, but realistic uncertainties in Fig.\ref{dTeff_Teff} (red triangles)
, without taking into-account the fact that the radius is unknown. 

We observe two regimes. First, lower than 800~K, there is no visible difference with the cases 
with and without observational uncertainties. This comes from the fact that the error is 
dominated by the lack of data because the maximum of the flux is near the maximum of MIRI LRS 
and so we lose a lot of flux.

Then, comparatively, for $T_\mathrm{eff} > 800$~K we observe an increase of the $\delta$T. 
This comes from the fact that the maximum of the flux is shifted toward the visible part of 
the spectrum, where NIRSpec has a poorer SNR. This, combined with the fact that the flux is 
high, i.e. the temperature is high, leads to $\delta$T exceeding 10~K.\\

This picture is not complete because the true data give the observed spectrum, that will 
be linked to the $F_{\lambda}$ in Eq.~\ref{eq-blackbody} through a "dilution factor" 
$\frac{R^2}{D^2}$ where $R$ is the radius of the planet and $D$ the distance between the planet 
and the observer. The radius is not known but can be inferred. One possibility is to use the 
spectra and evolution models combined with the aged of the systems as shown in 
\cite{Males2014} and \cite{Morzinski_2015}. This will give an estimation of the radius, mass and 
$T_\mathrm{eff}$. Using sparse ground-based data, \cite{Morzinski_2015} found errors around 
20~K and 0.02~$R_\mathrm{Jup}$. Even by using a more conservative error, we should be able to obtain 
similar or better result using JWST.

The $T_\mathrm{eff}$ will be obtained directly by JWST spectroscopy with a good uncertainty, 
probably similar or better than the current uncertainties derived from ground-based observations 
compared with models (stars in Fig.~\ref{dTeff_Teff}).

Being able to achieve this precision without using models will be crucial to obtain benchmark 
cases to improve modelling (self-consistent and retrieval). As the community is currently 
trying to deal with many unknowns to generate accurate spectra \citep{Baudino_2017} 
especially for young (or highly irradiated) objects, being able to fix one important 
parameter such as the $T_\mathrm{eff}$ will narrow the range of acceptable solutions to a more 
manageable set. If the differences between models (such as alkali and molecular far wing 
lineshape) are known, it will enable us to identify the best hypothesis in these models. This 
work will need to be done before any precise analysis of the JWST observation can be made and 
the direct-imaging ERS proposal assesses this properly as VHS~1256-1257~b will be observed 
using the modes used previously in this paper.

\section{Conclusions}
\label{sec-5}

We performed an analysis of the uncertainty of the estimated $T_\mathrm{eff}$ of exoplanets 
depending on the observed/defined wavelength range and we quantified the link between using a 
given wavelength range and the inferred $T_\mathrm{eff}$. This can be used to optimise the 
modelling of exoplanet atmospheres. Computing the spectrum over a large wavelength range is 
time consuming so modellers need to choose a maximum wavelength. We have shown here that a 
maximum wavelength greater than 70 $\mu$m is sufficient to have an intrinsic error on the 
$\delta T_\mathrm{eff} <$~1~K, for planets with an $T_\mathrm{eff}$ in the range 400--1800~K. 
If models use a shorter wavelength range then Fig.~\ref{Teffvsrange} can help to impose a 
modelling uncertainty on the $T_\mathrm{eff}$.\\

In cases where we will be able to obtain the full spectrum combining NIRSpec and MIRI we shown 
that we will obtain directly a good (similar or better than what we can currently achieve 
with SPHERE or GPI) characterisation of the $T_\mathrm{eff}$ (i.e. with $\delta$~T~$<100$~K). 
This kind of characterisation will be crucial to benchmark and update our models and 
approaches to analyse exoplanet direct-imaging observations. In the context of the JWST it 
will be mandatory to take into account the intrinsic modelling errors, as shown in this 
paper, for the $T_\mathrm{eff}$.\\

Obtaining an accurate independent evaluation of $T_\mathrm{eff}$ will lead us to improve our 
models (self-consistent and retrieval). To succeed to do so properly we will need to have a 
clear view of the differences and similarities between the models available in the 
literature. One of the methods to do so is to continue the benchmark process initiated by 
\cite{Baudino_2017}.

\section*{Acknowledgements}

J.L.B. acknowledges the support of the UK Science and Technology Facilities Council. 
This publication makes use of VOSA, developed under the Spanish Virtual Observatory project 
supported from the Spanish MICINN through grant AyA2011-24052. This research made use of 
Numpy: \cite{Oliphant_2006}, \cite{VanDerWalt_2011}. This research made use of Astropy,
\footnote{http://www.astropy.org} a community-developed core Python package for Astronomy 
\citep{AstropyCollaboration_2013, AstropyCollaboration_2018}. The authors thank the referee 
for the useful comments. 

%%%%%%%%%%%%%%%%%%%%%%%%%%%%%%%%%%%%%%%%%%%%%%%%%%



%\bibliographystyle{mnras}
\bibliographystyle{mnras}
\bibliography{bibtex}

%\begin{thebibliography}{99}
%\end{thebibliography}

%%%%%%%%%%%%%%%%%%%%%%%%%%%%%%%%%%%%%%%%%%%%%%%%%%

%%%%%%%%%%%%%%%%% APPENDICES %%%%%%%%%%%%%%%%%%%%%

\appendix

%%%%%%%%%%%%%%%%%%%%%%%%%%%%%%%%%%%%%%%%%%%%%%%%%%

% Don't change these lines
\bsp	% typesetting comment
\label{lastpage}
\end{document}